\documentclass{article}

\usepackage[all]{xy,xypic}
\usepackage{amsfonts,amssymb,amsmath,amsgen,amsopn,amsbsy,theorem,graphicx,epsfig}
\usepackage{eufrak,amscd,bezier,latexsym,mathrsfs,enumerate,multirow}
\usepackage[utf8]{inputenc}\usepackage[english]{babel}
\usepackage[dvipsnames]{xcolor}
\usepackage[pagewise]{lineno}
\usepackage{epstopdf}
\usepackage{tabularx}
\usepackage{relsize}
\usepackage{float}
\usepackage[numbers,sort&compress]{natbib}%
\usepackage{graphicx}
\allowdisplaybreaks
\title{The valuation of variance swaps under stochastic volatility, stochastic interest rate and full correlation structure}
\begin{document}

\maketitle


\author[T.R.N Roslan]{Teh Raihana Nazirah Roslan}

\author[W. Zhang]{Wenjun Zhang}

\author[J. Cao]{Jiling Cao}


\begin{abstract}
This paper considers the case of pricing discretely-sampled variance swaps under the class of equity-interest rate hybridization. Our modeling framework consists of the equity which follows the dynamics of the Heston stochastic volatility model, and the stochastic interest rate is driven by the Cox-Ingersoll-Ross (CIR) process with full correlation structure imposed among the state variables. This full correlation structure possess the limitation to have fully analytical pricing formula for hybrid models of variance swaps, due to the non-affinity property embedded in the model itself. We address this issue by obtaining an efficient semi-closed form pricing formula of variance swaps for an approximation of the hybrid model via the derivation of characteristic functions. Subsequently, we implement numerical experiments to evaluate the accuracy of our pricing formula. Our findings confirmed that the impact of the correlation between the underlying and the interest rate is significant for pricing discretely-sampled variance swaps.

\end{abstract}

\maketitle

\section{Introduction}

The study of finance largely concerns about the trade-off between risk and expected return. A significant source of risk in financial market is the uncertainty of the volatility of equity indices, where volatility is understood as the standard deviation of a financial instrument's return with a specific time horizon.
In late 1990s, Wall Street firms started trading volatility derivatives such as variance swaps. Since then, these derivatives have become a preferred route for many hedge fund managers to trade on market volatility. Due to the crucial role that volatility plays in making investment decisions, it is important for financial practitioners to understand the nature of the volatility variations. Research on volatility derivatives has been an active pursued topic in quantitative finance.

Researchers working in the field concerning volatility derivatives have been focusing on developing suitable methods for evaluating variance swaps.  Carr and Madan \cite{5} combined static replication using options with dynamic trading in futures to price and hedge certain volatility contracts without specifying the volatility process. The principal assumptions were continuous trading and continuous semi-martingale price processes for the future prices.  Demeterfi et al.~\cite{8} worked in the same area by proving that a variance swap could be reproduced via a portfolio of standard options. The requirements were continuity of exercise prices for the options and continuous sampling time for the variance swaps. One common feature shared among these researches was the assumption of continuous sampling time which was actually an simplification of the discrete sampling reality in financial markets. In fact, options of discretely-sampled variance swaps were mis-valued when the continuous sampling was used as approximation, and large inaccuracies occurred in certain sampling periods, as discussed in \cite{1},\cite{10},\cite{18},\cite{22}.

In addition to the above mentioned analytical approaches, some other authors also conducted researches using numerical approaches. Little and Pant \cite{18} explored the finite-difference method via dimension-reduction approach and obtained high efficiency and accuracy for discretely-sampled variance swaps.   Windcliff et al.~\cite{21} investigated the effects of employing the partial-integro differential equation on constant volatility, local volatility and jump diffusion-based volatility products.  An extension of the approach in \cite{18} was made by Zhu and Lian in \cite{22} through incorporating Heston two-factor stochastic volatility for pricing discretely-sampled variance swaps. Another recent study was conducted by Bernard and Cui \cite{1} on analytical and asymptotic results for discrete sampling variance swaps with three different stochastic volatility models. Their Cholesky decomposition technique exhibited significant simplification.  However, the  constant interest rate assumption  by the authors  did not reflect the real market phenomena.

One of the contemporary developments in the financial research was the emergence of hybrid models, which described interactions between different asset classes such as stock, interest rate and volatility. The main aim of these models was to provide customized alternatives for market practitioners and financial institutions, as well as reducing the associated risks of the underlying assets. Hybrid models could be generally categorized into two different types, namely hybrid models with full correlation and hybrid models with partial correlation among engaged underlyings. Literatures concerning hybrid models with partial correlation among asset classes appeared to dominate the field due to less complexity involved. Majority of the researchers focused on either inducing correlation between the stock and interest rate, or between the stock and the volatility. Grunbichler and Longstaff \cite{11} developed pricing model for options on variance based on the Heston stochastic volatility model. \cite{6} and \cite{13} stressed that correlation between equity and interest rate was crucial to ensure that the pricing activities were precise, especially for industrial practice. According to these authors, the correlation effects between equity and interest rate were more distinct compared to the correlation effects between interest rate and volatility. The hybrid models with full correlation among underlyings started to attract attention for their improved model capability. \cite{14} and \cite{20} compared their Heston-Hull-White hybrid model with the SZHW hybridization for pricing inflation dependent options and European options, respectively.

In this article we develop the modeling framework that extends the Heston stochastic volatility model by including the stochastic interest rate which follows the CIR process. Note that \cite{4} derived a semi-analytical pricing formula for partially correlated Heston-CIR hybrid model of discretely-sampled variance swaps. Their suggestion of imposing full correlation among state variables is considered in this work. Our focus is on the pricing of discrete sampling variance swaps with full correlation among equity, interest rate as well as volatility. Since the Heston-CIR model hybridization is not affine, we approach the pricing problem via the hybrid model approximation which fits in the class of affine diffusion models \cite{9,12}. The key ingredient involves the derivation of characteristic functions for two phases of partial differential equations and we obtain a semi-closed form  pricing formula for variance swaps. Numerical experiments are performed to evaluate the accuracy of the pricing formula.
\section{Specification of the variance swaps pricing model}

In this section we present a hybrid model which combines the Heston stochastic volatility model with the one-factor CIR stochastic interest rate model. Our model extends the work in \cite{4} by imposing full correlation among the underling asset, volatility and interest rate. Recently, \cite{17} proposed a model which was a combination of the multi-scale stochastic volatility model and the Hull-White interest rate model and showed that incorporation of the stochastic interest rate process into the stochastic volatility model gave better results compared with the constant interest rate case in any maturity.

\subsection{The Heston-CIR hybrid model}
Given $T>0$, let $\{S(t): 0\le t \le T\}$ be the stochastic process of some
  asset price with the time horizon $[0,T]$. The Heston-CIR hybrid
  model under the real world measure $\mathbb P$ is as follows
  \begin{equation}
  \left \{\label{Eq2.1}
  \begin{array}{ll}
  dS(t)=\mu S(t)dt+\sqrt{\nu(t)}S(t)dW_{1}(t), \quad 0\le t \le T,
  \\[0.5em]
  d\nu(t)=\kappa(\theta-\nu(t))dt+\sigma\sqrt{\nu(t)}dW_{2}(t),
  \quad 0\le t \le T,\\[0.5em]
  dr(t)=\alpha(\beta-r(t))dt +\eta \sqrt{r(t)} dW_{3}(t), \quad
  0\le t \le T,
  \end{array}
  \right.
  \end{equation}
  where $\{\nu(t): 0\le t \le T\}$ and $\{r(t): 0 \le t \le T\}$  are the stochastic instantaneous
  variance process and the stochastic instantaneous interest rate process, respectively. In the stochastic instantaneous variance process $\nu(t)$, the
  parameter $\theta$ is its long-term mean, $\kappa$ governs the speed of mean reversion
   and $\sigma$ is the volatility of the volatility.
  Similarly in the stochastic instantaneous variance process $r(t)$, $\beta$ is the interest rate term
  structure, $\alpha$ controls the mean-reverting speed and $\eta$ determines the volatility of the interest
  rate. In order to ensure that the square root processes in
  $\nu(t)$ and $r(t)$ are always positive, it is required that $2\kappa
  \theta \geq \sigma^2$ and $2\alpha \beta \geq \eta^2$ respectively,
  refer to \cite{7,15}.
  The correlation involved
  are given by
 $(dW_{1}(t), dW_{2}(t))=\rho_{12}dt=\rho_{21}dt,$
  $(dW_{1}(t), dW_{3}(t))=\rho_{13}dt=\rho_{31}dt,$
  and
$(dW_{2}(t), dW_{3}(t))=\rho_{23}dt=\rho_{32}dt,$
where $-1\le \rho_{ij} \le 1$ for all $i,j=1,2,3$.

  According to the Girsanov theorem, there exists a
  risk-neutral measure $\mathbb Q$ equivalent to
  the real world measure $\mathbb P$ such that under
  $\mathbb Q$ the Heston-CIR model can be  described as
  \begin{equation}
  \left \{\label{Eq2.2}
  \begin{array}{ll}
  dS(t)=r(t)S(t)dt+\sqrt{\nu(t)}S(t)d\widetilde{W}_{1}(t),
  \quad 0\le t \le T,\\[0.5em]
  d\nu(t)=\kappa^*
  (\theta^*-\nu(t))dt+\sigma\sqrt{\nu(t)}d\widetilde{W}_{2}(t),
  \quad 0\le t \le T,\\[0.5em]
  dr(t)=\alpha^*(\beta^*-r(t))dt +\eta \sqrt{r(t)} d\widetilde{W}_{3}(t),
  \quad 0\le t \le T,
  \end{array}
  \right.
  \end{equation}
  where the risk-neutral parameters are given as $\kappa^* = \kappa+\lambda_1$, $\theta^* =\frac{\kappa\theta}
  {\kappa+\lambda_1}$, $\alpha^*=\alpha+\lambda_2$ and $\beta^*=
  \frac{\alpha\beta}{\alpha+\lambda_2}$,  and the parameters
  $\lambda_1$ and $\lambda_2$ represent the premium prices of volatility and interest
  rate risk, respectively.
  The Brownian motion under $\mathbb Q$ is denoted by $\{\widetilde{W}_{i}(t): 0\le t \le T\}$
  ($1\le i\le 3$).

 Using the Cholesky decomposition,  we can re-write SDEs \eqref{Eq2.2} in terms of independent Brownian motions as
  \begin{equation}\label{Eq2.3}
  \left( \begin{array}{cc} \frac {dS(t)}{S(t)} \\[0.5em]
  d\nu(t)\\ dr(t)\\ \end{array} \right)
  = \mu^{\mathbb Q} dt
  +\Sigma \times L \times \left( \begin{array}{cc} d W^*_{1}(t)
  \\[0.5em]
  d W^*_{2}(t) \\[0.5em]
  d W^*_{3}(t)\end{array} \right),\;\;
  0\le t\le T,
  \end{equation}
  where
  \begin{equation*}
  \mu^{\mathbb Q}=\left( \begin{array}{cc} r(t) \\ \kappa^*(\theta^*-\nu(t))\\
  \alpha^*(\beta^*-r(t))\\ \end{array} \right), \quad
  \Sigma=\left(\begin{array}{ccc} \sqrt{\nu(t)} & 0 & 0 \\ 0 &
  \sigma\sqrt{\nu(t)} & 0 \\ 0 & 0 & \eta \sqrt{r(t)} \end{array} \right),\\
  \end{equation*}
  \mbox{ and }
  \begin{equation*}
  L =\left(\begin{array}{ccc} 1 & 0 & 0 \\ \rho_{12} & \sqrt{1-\rho_{12}^2}
  & 0 \\ \rho_{13} & \dfrac{\rho_{23}-\rho_{13}\rho_{12}}{\sqrt{1-\rho_{12}^2}} & \sqrt{{1-\rho_{13}^2}-\left(\dfrac{\rho_{23}-\rho_{13}\rho_{12}}{\sqrt{1-\rho_{12}^2}}\right)^2} \end{array} \right)
  \end{equation*}
  such that
  \begin{equation*}
  L L^\top=\left(\begin{array}{ccc} 1 & \rho_{12} & \rho_{13} \\ \rho_{21} & 1 & \rho_{23} \\ \rho_{31} & \rho_{32} & 1 \end{array} \right).
  \end{equation*}
  Here, $W^*_{1}(t)$, $W^*_{2}(t)$ and $W^*_{3}(t)$ are three Brownian motions under
  $\mathbb Q$ such that $dW^*_{1}(t)$, $dW^*_{2}(t)$ and $dW^*_{3}(t)$ are mutually
  independent and satisfy the following relation
  \begin{equation*}
  \left( \begin{array}{cc} d\widetilde{W}_{1}(t)
  \\[0.5em]
  d\widetilde{W}_{2}(t) \\[0.5em]
  d\widetilde{W}_{3}(t)\end{array} \right)
  = L \times
  \left( \begin{array}{cc} dW^*_{1}(t)
  \\[0.5em]
  dW^*_{2}(t) \\[0.5em]
  dW^*_{3}(t) \end{array} \right), \quad 0\le t\le T.
  \end{equation*}

\subsection{Valuation of variance swaps}

Variance swaps were first launched in 1990s due to the breakthrough of volatility derivatives in the market.
Since the payment of a variance swap is only made in a single payment at maturity, it is defined as a forward contract on the future realized variance of
  the returns of the underlying asset. 
  Suppose that the underlying asset $S(t)$ is observed  $N$ times during the contract period and $t_{j}$ denotes the \emph{j}-th observation time, then a typical formula for the measure of realized variance, denoted as $RV$, is given by
  \begin{equation}
  RV=\frac{AF}{N}\sum_{j=1}^N \left(\frac{S(t_{j})-S(t_{j-1})}{S(t_{j-1})} \right)^2 \times 100^2, \label{Def1}
  \end{equation}
  where $AF$ is the annualized factor which converts the above expression to annualized variance points depending on the sampling frequency.
  The measure of realized variance requires sampling the underlying price path discretely, usually at the
  end of each business day, so $AF$ is 252 in such case. If the sampling frequency is every month or every week, then $AF$ will be $12$ and $52$ respectively.

  At maturity time $T$, a variance swap rate is $V(T)=
  (RV-K)\times L$, where $K$ is the annualized delivery price for the
  variance swap and $L$ is the notional amount of the swap in dollars.
  In the risk-neutral world, the value of a variance swap with stochastic interest rate at
  time $t$ is the expected present value of its future payoff amount, that is,
  $V(t)= \mathbb{E}_t^{\mathbb Q}\left[e^{-\int_{t}^{T}r(s)ds}(RV-K)\times L
  \right]$. This value should be zero at $t=0$, since it is defined in the class of forward contracts.
  The above expectation calculation involves the joint distribution of the interest rate and the future payoff, so it is complicated to evaluate. Thus, it would be more convenient to use the bond price as the numeraire,
since the price of a $T$-maturity zero-coupon bond at $t=0$ is given by $\mathbb{E}_0^{\mathbb Q} \left[ e^{-\int_{0}^{T}r(s)ds}\right]$.
  We can determine the value of $K$ by changing $\mathbb Q$ to the $T$-forward measure ${\mathbb Q}^T$.
  It follows that
  \begin{equation}
  \mathbb{E}_0^{\mathbb Q} \left[e^{-\int_{0}^{T}r(s)ds}(RV-K)\times L \right]=\mathbb{E}_0^{\mathbb Q} \left[ e^{-\int_{0}^{T}r(s)ds}\right]\mathbb{E}_0^{T}(RV-K)\times L,
  \end{equation}
  where $\mathbb{E}_0^{T}(\cdot)$ denotes the expectation with respect
  to ${\mathbb Q}^T$ at $t=0$. Thus, the fair delivery price of the
  variance swap is given by $K=\mathbb{E}_0^T[RV]$.

\subsection{Variance swaps dynamics under the $T$-forward measure}

Under the $T$-forward measure, the valuation of the fair delivery
  price for a variance swap is reduced to calculating the $N$ expectations expressed in the form of
  \begin{equation}
  \mathbb{E}_0^T \left[\left(\frac{S(t_{j})-S(t_{j-1})}{S(t_{j-1})}\right)^2 \right]
  \label{MainExpectation}
  \end{equation}
  for $t_0=0$, some fixed equal time period $\Delta t$ and $N$ different tenors\ $t_{j}=j\Delta t$  $(j=1,\cdots,N)$.
  It is important to note that we have to consider two cases $j=1$ and $j>1$ separately. For the case $j=1$, we have $t_{j-1}
  =0$ and $S(t_{j-1})=S(0)$ is a known value, instead of an
  unknown value of $S(t_{j-1})$ for any other cases with $j>1$. In the process of finding this expectation, $j$, unless otherwise
  stated, is regarded as a constant. Hence both $t_{j}$ and $t_{j-1}$ are
  regarded as known constants.

  Based on the tower property of conditional expectations, the calculation of expectation (\ref{MainExpectation}) can be separated into
  two phases in the following form
  \begin{equation}\label{tower}
  \begin{array}{ll}
  \mathbb{E}_0^T \left[\left(\frac{S(t_{j})}{S(t_{j-1})}-1\right)^2 \right]
  =\mathbb{E}_0^T \left[\mathbb{E}_{t_{j-1}}^T \left[\left(\frac{S(t_{j})}{S(t_{j-1})}-1\right)^2\right]\right].
  \end{array}
  \end{equation}
  We denote the term $\mathbb{E}_{t_{j-1}}^T \left[\left(\frac{S(t_{j})}{S(t_{j-1})}-1\right)^2
  \right]$ by $G_j(\nu(t_{j-1}) ,r(t_{j-1}))$ for notational convenience. Then,
   in the first phase, the computation involved is to find
   $G_j(\nu(t_{j-1}) ,r(t_{j-1}))$,
  and in the second phase, we need to compute
  \begin{equation}\label{tower2}
  \mathbb{E}_0^T \left[ G_j(\nu(t_{j-1}) ,r(t_{j-1})) \right].
  \end{equation}

  To this purpose, we implement the measure change from risk neutral measure $\mathbb Q$ to
  the $T$-forward measure ${\mathbb Q}^T$. Note that the numeraire
  under $\mathbb Q$ is $N_{1,t} = e^{\int_{0}^{t}r(s)ds}$, whereas the
  numeraire under ${\mathbb Q}^{T}$ is $N_{2,t}= A(t,T)e^{-B(t,T)r(t)}$, refer to \cite{3}.
  Implementation of the Radon-Nikodym derivative for these two
  numeraires gives the dynamics  for (\ref{Eq2.3}) under $\mathbb{Q}^T$
  as follows
    \begin{eqnarray}\label{ForwardMeasureSDE}
    \left( \begin{array}{cc} \frac {dS(t)}{S(t)} \\ d\nu(t)\\ dr(t)\\ \end{array} \right)
&  = & \Sigma \times L \times \left( \begin{array}{cc} dW^*_{1}(t)\\dW^*_{2}(t)\\dW^*_{3}(t)\end{array} \right)
  \\
&&+  \left( \begin{array}{cc} r(t)-\rho_{13}B(t,T)\eta\sqrt{r(t)}\sqrt{\nu(t)} \\ \kappa^*(\theta^*-\nu(t))-\rho_{23}\sigma B(t,T) \eta \sqrt{r(t)}\sqrt{\nu(t)}\\ \alpha^*\beta^*-(\alpha^*+B(t,T) \eta^2)r(t)\\ \end{array} \right) dt, \nonumber
  \end{eqnarray}
  where
  \begin{equation*}
  \begin{array}{ll}
  B(t,T)=\dfrac{2\left(e^{(T-t)\sqrt{(\alpha^*)^2+2\eta^2}}-1 \right)}{2\sqrt{(\alpha^*)^2 +2\eta^2}+
  \left(\alpha^*+\sqrt{(\alpha^*)^2+2\eta^2}\right)
  \left(e^{(T-t)\sqrt{(\alpha^*)^2+2\eta^2}}-1\right)}.
  \end{array}
  \end{equation*}
We present further details regarding the change of  measure in Appendix A.

\section{Solution techniques for pricing variance swaps}

\subsection{Solution for the first phase}

In order to find the term $G_j(\nu(t_{j-1}) ,r(t_{j-1}))$,
we consider a contingent claim denoted by $U_j(S(t),\nu(t),r(t),t)$ for $t\in [t_{j-1},t_j]$.
The  contingent claim  has a European-style payoff function at expiry $t_{j}$ denoted by
  \begin{equation}
   H_j(S)=\left(\frac{S}{S(t_{j-1})}-1\right)^2.
  \end{equation}
  Applying standard techniques in the general asset valuation
  theory, the PDE for $U_j$ over $[t_{j-1}, t_j]$ can be obtained as
  \begin{equation}\label{Eq1}
  \begin{array}{ll}
  \dfrac{\partial U_j}{\partial t}+\dfrac{1}{2}\nu S^2 \dfrac{\partial^2 U_j}{\partial S^2} + \dfrac{1}{2}\sigma^2\nu\dfrac{\partial^2 U_j}
  {\partial \nu^2} + \dfrac{1}{2}\eta^2 r \dfrac{\partial^2 U_j}
  {\partial r^2}+ \rho_{12}\sigma \nu S \dfrac{\partial^2 U_j}
  {\partial S\partial \nu}\\[0.8em] +\left(rS-\rho_{13}B(t,T)\eta\sqrt{r(t)}\sqrt{\nu(t)}S \right)\dfrac{\partial U_j}{\partial S}
  + \left(\alpha^*\beta^*-(\alpha^*+B(t,T)\eta^2)r\right) \dfrac{\partial U_j}{\partial r}
  \\[0.8em]
  +\left(\kappa^*(\theta^*-\nu)-\rho_{23}\sigma B(t,T)\eta\sqrt{r(t)}\sqrt{\nu(t)}\right) \dfrac{\partial U_j}{\partial \nu} +\rho_{23}\sigma \eta \sqrt{\nu(t)}\sqrt{r(t)}\dfrac{\partial^2 U_j}
  {\partial \nu \partial r}\\[0.8em]+\rho_{13}\eta \sqrt{\nu(t)}\sqrt{r(t)}S \dfrac{\partial^2 U_j}
  {\partial S\partial r}=0
  \end{array}
  \end{equation}
  with the terminal condition
  \begin{equation*}
  U_j(S,\nu,r,t_{j})=H_j(S).
  \end{equation*}
For notational convenience, we omit the subscript $j$, replace the expiry $t_j$ as $T$ and let
$\tau=T-t$ and $x=\ln{S}$,
 then \eqref{Eq1}
 is transformed into
 \begin{equation}
 \label{Eq2}
 \left \{
 \begin{array}{ll}
 \dfrac{\partial U}{\partial \tau}=\dfrac{1}{2} \nu \dfrac{\partial^2 U}{\partial
 x^2}+\dfrac{1}{2}\sigma^2 \nu\dfrac{\partial^2 U}{\partial \nu^2}+ \dfrac{1}{2}\eta^2 r \dfrac{\partial^2 U}{\partial r^2}+\rho_{12}\sigma \nu \dfrac{\partial^2 U}{\partial x\partial
 \nu}\\[0.8em]+\left(r-\rho_{13}B(T-\tau,T)\eta \sqrt{\nu(T-\tau)}\sqrt{r(T-\tau)}-\dfrac{1}{2}\nu \right)\dfrac{\partial
 U}{\partial
 x}\\[0.8em]+\left(\kappa^*\left(\theta^*-\nu \right)-\rho_{23}\sigma B(T-\tau,T)\eta\sqrt{\nu(T-\tau)}\sqrt{r(T-\tau)}\right)\dfrac{\partial U}{\partial
 \nu}\\[0.8em]+\left(\alpha^*\beta^*-(\alpha^*+B(T-\tau,T)\eta^2)r \right)\dfrac{\partial U}{\partial
 r}+\rho_{13}\eta \sqrt{\nu(T-\tau)}\sqrt{r(T-\tau)} \dfrac{\partial^2 U}{\partial x\partial
 r}\\[0.8em]+\rho_{23}\sigma \eta\sqrt{\nu(T-\tau)} \sqrt{r(T-\tau)}\dfrac{\partial^2 U}{\partial \nu \partial
 r}.\\[0.8em]
 U(x,\nu,r,0)=H(e^x).
 \end{array}
 \right.
 \end{equation}

 Next, we perform the generalized Fourier transform with respect to $x$ to find the solution of this PDE (refer to \cite{19}.
 As a result, the transformed PDE system of  $\widetilde{U}(\omega,\nu,r,\tau)=\mathcal{F}[U(x,\nu,r,\tau)]$ is
 \begin{equation}
  \label{eqn3.5}
 \left \{
 \begin{array}{ll}
 \dfrac{\partial \widetilde{U}}{\partial
 \tau}=\dfrac{1}{2}\sigma^2 \nu\dfrac{\partial^2 \widetilde{U}}{\partial \nu^2}
 +\dfrac{1}{2}\eta^2 r \dfrac{\partial^2 \widetilde{U}}{\partial r^2}\\[0.8em]+\left(\kappa^*\theta^* +(\rho_{12}\sigma\omega i-\kappa^*)\nu -\rho_{23}\sigma B(T-\tau,T)\eta\sqrt{\nu(T-\tau)}\sqrt{r(T-\tau)} \right)\dfrac{\partial
 \widetilde{U}}{\partial \nu} \\[0.8em]+\left(\alpha^*\beta^*-(\alpha^*+B(T-\tau,T)\eta^2)r +\rho_{13}\eta \sqrt{\nu(T-\tau)}\sqrt{r(T-\tau)} \omega i\right)\dfrac{\partial\widetilde{U}}{\partial r}\\[0.8em]+
 \rho_{23}\sigma \eta\sqrt{\nu(T-\tau)} \sqrt{r(T-\tau)}\dfrac{\partial^2 \widetilde{U}}{\partial \nu \partial
 r}\\[0.8em]+\left(-\dfrac{1}{2}(\omega i+\omega^2) \nu+r \omega i -\rho_{13}B(T-\tau,T)\eta \sqrt{\nu(T-\tau)} \sqrt{r(T-\tau)} \omega i \right)\widetilde{U},\\[0.8em]
 \widetilde{U}(\omega,\nu,r,0)=\mathcal{F}[H(e^x)],\\
 \end{array}
 \right.
 \end{equation}
where $i=\sqrt{-1}$ and $\omega$ is the Fourier transform variable.
 In order to solve the above PDE system, we adopt Heston's assumption in \cite{16} that the PDE solution has an affine form  as follows
 \begin{equation}
  \label{eqn3.6}
\widetilde{U}(\omega,\nu,r,\tau)=e^{C(\omega,\tau)+D(\omega,\tau)\nu +E(\omega,\tau)r}\widetilde{U}(\omega,\nu,r,0).
\end{equation}
 We can then obtain three ordinary differential equations by substituting the above function form \eqref{eqn3.6} into the PDE system \eqref{eqn3.5} as
  \begin{equation}
 \left \{
 \begin{array}{ll}
 \dfrac{dD}{d\tau}=\dfrac{1}{2}\sigma^2D^2+(\rho_{12}\omega\sigma
 i-\kappa^*)D-\dfrac{1}{2}\left(\omega^2+\omega i \right),\\[0.8em]
 \dfrac{dE}{d\tau}=\dfrac{1}{2} \eta^2 E^2 -(\alpha^*+B(T-\tau,T)\eta^2) E +\omega i, \\[0.8em]
 \dfrac{dC}{d\tau}=\kappa^*\theta^* D+ \alpha^* \beta^* E -\rho_{13}\eta \sqrt{\nu(T-\tau)}\sqrt{r(T-\tau)}\omega i B(T-\tau,T)\\[0.8em]\quad \quad \quad+\rho_{13}\eta \sqrt{\nu(T-\tau)}\sqrt{r(T-\tau)}\omega i E\\[0.8em]
 \quad \quad \quad -\rho_{23}\sigma\eta \sqrt{\nu(T-\tau)}\sqrt{r(T-\tau)}DB(T-\tau,T)\\[0.8em]\quad \quad \quad+\rho_{23}\eta \sigma \sqrt{\nu(T-\tau)}\sqrt{r(T-\tau)}DE,\\
 \end{array}
 \right.
 \end{equation}
 with the initial conditions
 \begin{equation*}
 C(\omega,0)=0, \quad D(\omega,0)=0, \quad E(\omega,0)=0.
 \end{equation*}
 Note that only the function $D$ has analytical form as
  \begin{equation*}
  \begin{array}{ll}
  D(\tau)=\dfrac{a+b}{\sigma^2}\dfrac{1-e^{b\tau}}
  {1-ge^{b\tau}}, \quad
  a=\kappa^*-\rho_{12}\sigma\omega i, \quad \\[0.8em]
  b=\sqrt{a^2+\sigma^2(\omega^2+\omega i)}, \quad g=\dfrac{a+b}{a-b}.
  \end{array}
  \end{equation*}
The approximate solutions of the functions $E$ and $C$ can be found by numerical integrations using standard mathematical software package, e.g., Matlab. The algorithm of evaluating the functions $E$ and $C$ is given in Appendix B.

 Since the Fourier transform variable $\omega$ appears
 as a parameter in functions $D$, $C$ and $E$, the inverse Fourier transform is conducted to retrieve
 the solution as in its initial setup
 \begin{equation*}
 \begin{array}{ll}
 U(x,\nu,r,\tau)=\mathcal{F}^{-1}\left[\widetilde{U}(\omega,\nu,r,\tau)\right]\\[0.8em]
 \quad \quad \quad \quad \quad =\mathcal{F}^{-1}\left[e^{C(\omega, \tau)+D(\omega,
 \tau)\nu + E(\omega,
 \tau)r}\mathcal{F}[H(e^x)]\right].
 \end{array}
 \end{equation*}
In \cite{2} the generalized Fourier transform  $\hat{f}$ of a function $f$ is defined to be $$\hat{f}(\omega)=\mathcal{F}[f(x)]=\int^{\infty}_{-\infty}f(x)e^{-i\omega x}dx.$$
  The function $f$ can be derived from $\hat{f}$ via the generalized inverse Fourier transform
   $$f(x)=\mathcal{F}^{-1}[\hat{f}(\omega)]=\frac{1}{2\pi}\int^{\infty}_{-\infty}\hat{f}(\omega)e^{i\omega x}d\omega.$$
 Note that the Fourier transformation of the function $e^{i\xi x}$ is
  \begin{equation*}
  \mathcal{F}[e^{i\xi x}]=2\pi\delta_{\xi}(\omega),\\
  \end{equation*}
  where $\xi$ is any complex number and
  $\delta_{\xi}(\omega)$ is the generalized delta function
  satisfying
  \begin{equation*}
  \int_{-\infty}^{\infty}\delta_{\xi}(x)\Phi(x)dx=\Phi(\xi).\\
  \end{equation*}
For notational convenience, let $I=S(t_{j-1})$.  Conducting the generalized Fourier transform for the payoff $H(e^x)=(\frac{e^x}{I}-1)^2$ with respect to $x$ gives
  \begin{equation}
  \mathcal{F}\left[\left(\frac{e^x}{I}-1\right)^2\right]
  =2\pi\left(\frac{\delta_{-2i}(\omega)}{I^2}-2\frac{\delta_{-i}
  (\omega)}{I}+\delta_{0}(\omega)\right).
  \end{equation}
  As a result, the solution of the PDE (\ref{Eq1}) is derived as follows
  \begin{equation}\label{solwithpayoff}
  \begin{array}{ll}
  U_j(S,\nu,r,\tau)=\mathcal{F}^{-1}\left[e^{C(\omega, \tau)+D
  (\omega, \tau)\nu +E(\omega, \tau)r} 2\pi \left(\dfrac{\delta_{-2i}(\omega)}{I^2}-2\dfrac{\delta_{-i}(\omega)}{I}
  +\delta_{0}(\omega)\right)\right]\\[0.8em]
  \quad\quad\quad\quad\quad\quad=\dfrac{e^{2x}}{I^2}e^{\widetilde{C}(\tau)+\widetilde{D}(\tau)
  \nu +\widetilde{E}(\tau)r}-\dfrac{2e^{x}}{I}e^{\widehat{C}(\tau)+\widehat{E}(\tau)r}+1
  \\[0.8em]
  \quad\quad\quad\quad\quad\quad=\dfrac{S^2}{I^2}e^{\widetilde{C}(\tau)+\widetilde{D}(\tau)
  \nu +\widetilde{E}(\tau)r}-\dfrac{2S}{I}e^{\widehat{C}(\tau)+\widehat{E}(\tau)r}+1,
  \end{array}
  \end{equation}
  where $t_{j-1}\leq t\leq t_{j}$ and $\tau= t_j-t$. We denote $\widetilde{C}
  (\tau)$, $\widetilde{D}(\tau)$ and $\widetilde{E}(\tau)$ as $C(-2i,\tau)$, $D(-2i,\tau)$ and $E(-2i,\tau)$
  respectively.
    In addition, $\widehat{C}(\tau)$ and $\widehat{E}(\tau)$ are the notations for
  $C(-i,\tau)$ and $E(-i,\tau)$ respectively. Note that $D(-i,\tau)=0$.

\subsection{Solution for the second phase}
In this subsection, we  continue to carry out the second phase in finding out the expectation $\mathbb{E}_0^T[ G_j(\nu(t_{j-1}) ,r(t_{j-1}))]$.
Following \eqref{solwithpayoff} and letting $\tau=\Delta t$ in $U_j(S,\nu,r,\tau)$, we obtain the inner expectation  $G_j(\nu(t_{j-1}) ,r(t_{j-1}))$ as
  \begin{eqnarray}
  &&G_j(\nu(t_{j-1}) ,r(t_{j-1}))\nonumber \\
  &=&U_j(S,\nu,r,\Delta t) \\
  &=&e^{\widetilde{C}(\Delta
  t)+\widetilde{D}(\Delta t)\nu(t_{j-1})+\widetilde{E}(\Delta t)r(t_{j-1})} -2 e^{\widehat{C}(\Delta t)+\widehat{E}(\Delta t)r(t_{j-1})}+1. \nonumber
  \end{eqnarray}
  The outer expectation, $\mathbb{E}_0^T[ G_j(\nu(t_{j-1}) ,r(t_{j-1}))]$, is represented by
  \begin{eqnarray}
  \label{funcDoubleExpectation}
    &&G_j(\nu(0),r(0))\nonumber \\
  =&&\mathbb{E}_0^T[ G_j(\nu(t_{j-1}), r(t_{j-1}))] \nonumber  \\[0.8em]
  =&&  \mathbb{E}_0^T\left[
  e^{\widetilde{C}(\Delta
  t)+\widetilde{D}(\Delta t)\nu(t_{j-1})+\widetilde{E}(\Delta t)r(t_{j-1})} -2 e^{\widehat{C}(\Delta t)+\widehat{E}(\Delta t)r(t_{j-1})}+1 \right]\\[0.8em]
  =&&
  \mathbb{E}_0^T\left[
  e^{\widetilde{C}(\Delta
  t)+\widetilde{D}(\Delta t)\nu(t_{j-1})+\widetilde{E}(\Delta t)r(t_{j-1})}\right]
  - 2 \mathbb{E}_0^T \left[ e^{\widehat{C}(\Delta t)+\widehat{E}(\Delta t) r(t_{j-1}) } \right] +1 \nonumber \\
  = &&   e^{\widetilde{C}(\Delta
  t)} \cdot \mathbb{E}_0^T\left[ e^{\widetilde{D}(\Delta t)\nu(t_{j-1})+\widetilde{E}(\Delta t)r(t_{j-1})}\right]-2e^{\widehat{C}(\Delta t)} \cdot \mathbb{E}_0^T\left[ e^{ \widehat{E}(\Delta t) r(t_{j-1})} \right] +1.\nonumber
  \end{eqnarray}
  \medskip
In Appendix C, we show in more details how to derive approximate solutions for $\mathbb{E}_0^T\left[ e^{\widetilde{D}(\Delta t)\nu(t_{j-1})+\widetilde{E}(\Delta t)r(t_{j-1})}\right]$ and $\mathbb{E}_0^T\left[ e^{ \widehat{E}(\Delta t) r(t_{j-1})} \right]$ by using approximations of normally distributed random variable and its characteristic function.
\subsection{Delivery price of a variance swap}

In the previous two subsections, we demonstrate our solution techniques for pricing variance swaps by separating them into phases.
 However, as mentioned in Section 2.3, we have to consider two cases $j=1$ and $j>1$ separately. The case $j>1$ follows directly the
 expression in (\ref{funcDoubleExpectation}). For the case of $j=1$, we use the method described in Section 3.1 to obtain
  \begin{eqnarray*}
   G(\nu(0),r(0))&=&\mathbb{E}_0^T\left[\left(\dfrac{S(t_1)}{S(0)}-1\right)^2\right]\\
  &=&e^{\widetilde{C}(\Delta
  t)+\widetilde{D}(\Delta t)\nu(0)+\widetilde{E}(\Delta t)r(0)} -2 e^{\widehat{C}(\Delta t)+\widehat{E}(\Delta t)r(0)}+1.
  \end{eqnarray*}
The summation for the whole period from $j=1$ to $j=N$ gives the fair delivery price of a variance swap as
  \begin{equation}
  \label{finalresult}
  K=\mathbb{E}_0^T[RV] =\dfrac{100^2}{T}\left(G(\nu(0),r(0))+ \sum_{j=2}^{N}
  G_j(\nu(0),r(0)) \right).
  \end{equation}

\section{Numerical results}
\label{numerical}

In order to analyze the performance of our approximation formula \eqref{finalresult} for evaluating prices of variance swaps as described in the previous section, we conduct some numerical simulations. Comparisons are made with the Monte Carlo (MC) simulation which resembles the real market. In addition, we also investigate the impact of full correlation setting among the state variables in our model.

  Table 1 shows the set of parameters that we use for
  all the numerical experiments, unless otherwise stated.
  \begin{table}[htb]
  \caption{\label{ParaHeston} Model parameters of the Heston-CIR hybrid model.}
  \renewcommand\tabcolsep{1pt}
  \begin{tabularx}{\linewidth}{@{\extracolsep{\fill}}cccc|cccc|ccc c|c}
  \hline
  $S_0$ &  $\rho_{12}$ &$\rho_{13}$&$\rho_{23}$& $V_0$ & $\theta^*$  & $\kappa^*$ & $\sigma  $ & $r_0$
  & $\alpha^*$ & $\beta^*$ &  $\eta$ &  $T$ \\
  \hline
  1 & -0.4  & 0.5&0.5&0.05 & 0.05 & 2  &  0.1  & 0.05 &
  1.2 & 0.05 & 0.01 & 1 \\
  \hline
  \end{tabularx}
  \end{table}

\subsection{Comparison with MC simulation}

The MC simulation is a widely utilized numerical tool for the basis of conducting computations involving random variables.
We perform our MC simulation in this paper using the Euler-Maruyama scheme with $200,000$ sample paths.

  We present the comparison results between numerical implementation of the formula (\ref{finalresult}) with the MC simulation in Figure 1 and in Table 2. All values for the fair delivery prices are measured in variance points. 
  It could be seen in Figure 1 that our approximation formula matches the MC simulation very well. To gain some insight of the relative
  difference between our formula and the MC simulation, we compare their relative percentage error. By taking $N=52$ which is the weekly sampling frequency
   and $200,000$ paths, we discover that the error is $0.07\%$, with further reduction of the error as path numbers increase to $500,000$. Furthermore, even for small sampling frequency such as the quarterly sampling frequency when $N=4$, our formula can be executed in just $0.49$ seconds compared to  $27.7$ seconds needed by the MC simulation. These findings verify the accuracy and efficiency of our formula.
\begin{figure}[htb]
		\includegraphics{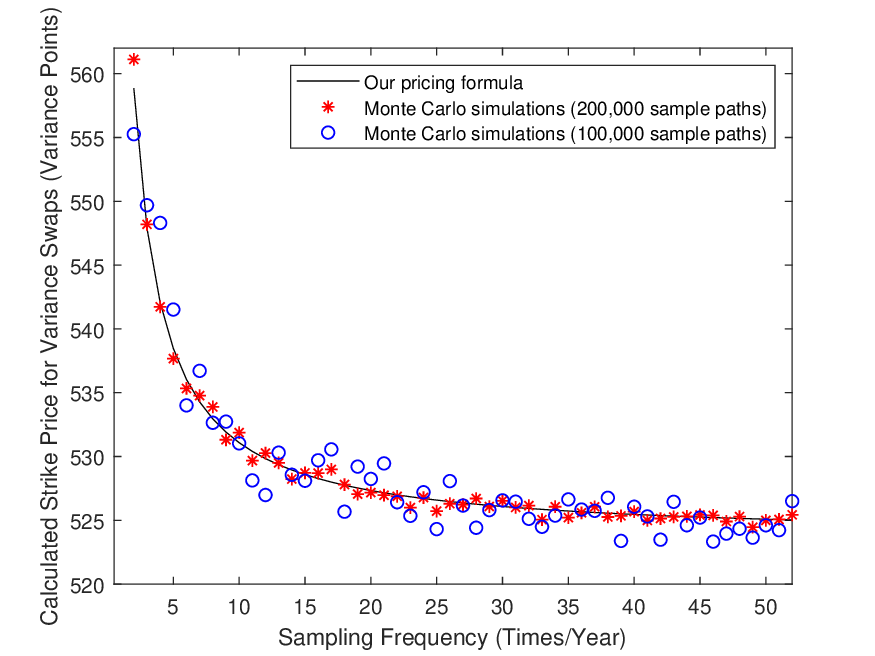}	
\vspace*{8pt}
\caption{Comparison of the delivery price of variance swaps between the formula (\ref{finalresult}) and the MC simulation.}
\end{figure}

\begin{table}[htb]
\begin{center}
\caption{Comparing variance swap prices between the results of formula (\ref{finalresult})  and the MC simulation.}
{\begin{tabular}{@{}lcccccc}
\hline
 Frequency &  \parbox[t]{2.3cm}{Formula result} & \parbox[t]{2.5cm}{\centering MC simulation \\(100,000 sample paths)}& \parbox[t]{2.5cm}{\centering Relative error \\between pricing formula and MC simulations (100,000 paths)} &
  \parbox[t]{2.5cm}{\centering MC simulation \\(200,000 sample paths)}
 & \parbox[t]{2.5cm}{\centering Relative error \\between pricing formula and MC simulations (200,000 paths)}\\
\hline
  N=4  &  542.06 &541.38 &0.125\% & 541.73 &0.061\% \\
  N=12  &  529.84   & 529.03& 0.153
  
  \%& 530.27&0.081\% \\
  N=26  &  526.47   & 527.05&0.110\%&526.30&0.032\% \\
  N=52  &  525.03   & 525.88&0.162\%&525.43&0.076\% \\
  N=252  &  523.89  & 524.42&0.101\%&524.10&0.040\%  \\
  \hline
\end{tabular}}
\label{comparisonresult}
\end{center}
\end{table}
\subsection{Impact of correlation among asset classes}
Next, we investigate the impact of the correlation coefficient between the
interest rate and the underlying $\rho_{13}$ and the correlation coefficient between the
interest rate and the volatility $\rho_{23}$, respectively.
The impact
of the correlation between the interest rate and the underlying is shown
in Figure 2.
In the figure we can see that the values of variance
swaps are increasing corresponding to the increase in the correlation values of $\rho_{13}$.
The difference of the variance swap rates goes up to 5 variance points for largely different correlation coefficient values of $\rho_{13}$.
This is very crucial since a relative difference of $2\%$
might produce considerable error.
However, it is also observed that the impact of the correlation
coefficient $\rho_{13}$ becomes less apparent as the sampling times increase.
\begin{figure}[htb]
\centerline{\includegraphics[width=4.0in]{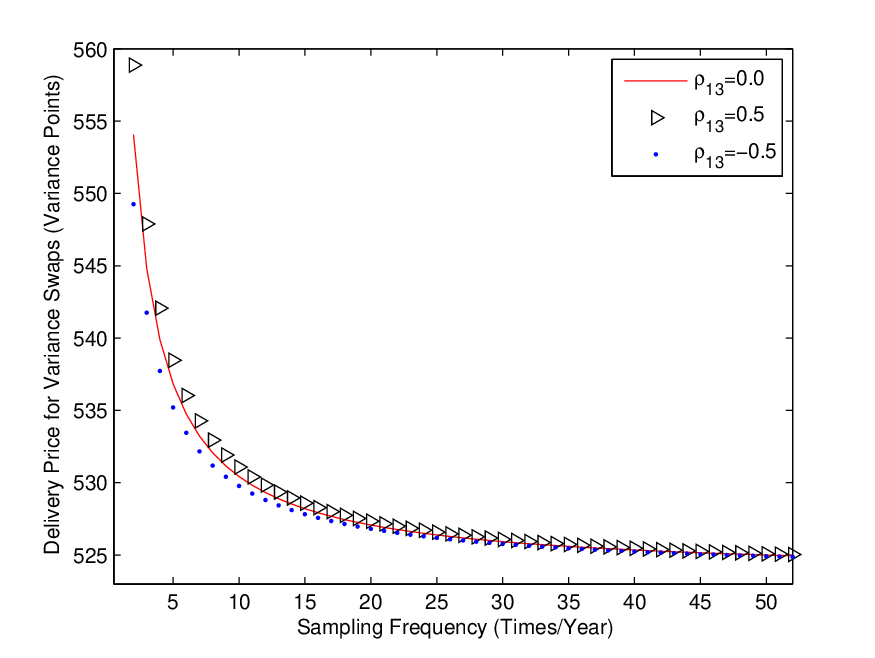}}
\vspace*{8pt}
\caption{Impact of different $\rho_{13}$ values on delivery price of variance swaps in the Heston-CIR hybrid model.}
\end{figure}
\begin{figure}[htb]
\centerline{\includegraphics[width=4.0in]{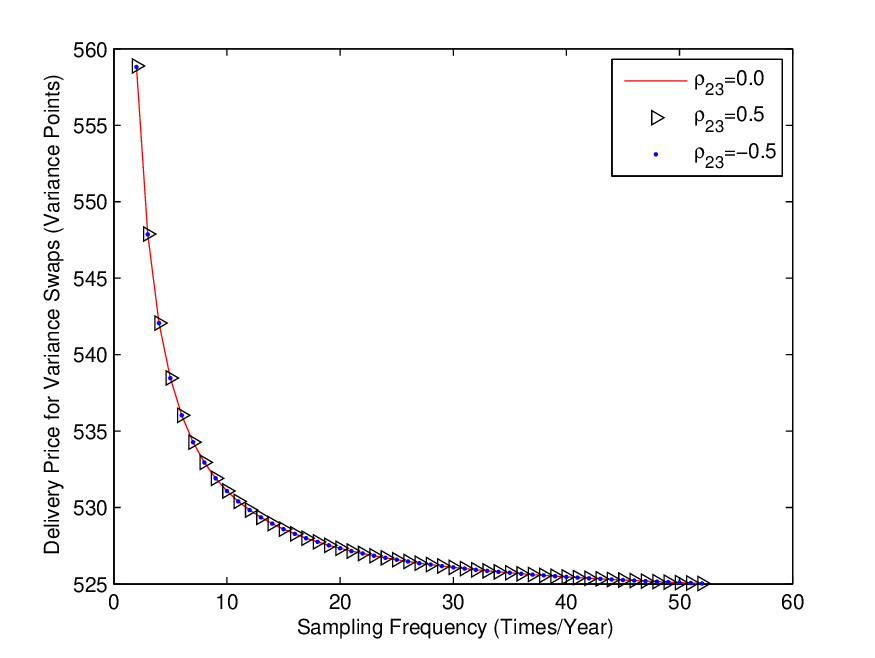}}
\vspace*{8pt}
\caption{Impact of different $\rho_{23}$ values on delivery price of variance swaps in the Heston-CIR hybrid model.}
\end{figure}
The effects of the correlation coefficient between the interest rate and the volatility are displayed in Figure 3.
In contrast to the significant correlation effects of $\rho_{13}$ in Figure 2, smaller impact of $\rho_{23}$ is observed. In
fact, the variance swap rates for three different values of $\rho_{23}$ are almost the same.
For example, for $N=12$ which is the monthly sampling frequency, the delivery price is $529.834$ for $\rho_{23}=0$, with only a slight increase
to $529.836$ for $\rho_{23}=0.5$, and a slight decrease to $529.833$ for $\rho_{23}=-0.5$. Figure 3 also displays the same trend of diminishing
impact of the correlation as the number of sampling periods increases.

\section{Conclusion}
This paper studies the evaluation of discretely-sampled variance swap rates in  the Heston-CIR hybrid model of  stochastic volatility and stochastic interest rate. This work extends the model framework considered in \cite{4} by imposing the full  correlation structure among the state variables. The proposed hybrid model is not affine, and we derive a semi-closed form approximation formula for the fair delivery price of variance swaps.
We consider the numerical implementation of our pricing formula which is validated to be fast and accurate via comparisons with the Monte Carlo simulation.
This pricing formula could be a useful tool for the purpose of model calibration to market quotation prices.
Our pricing model which offers the flexibility to correlate the underlying with both the volatility and the interest rate is a more realistic
model with practical importance for pricing and hedging. In fact, our numerical experiments confirm that
the impact of the correlation coefficient between the underlying and interest rates is very crucial, as it becomes more apparent for
larger correlation values. The pricing approach in our paper can be applied to other stochastic interest rate and stochastic volatility models, such as the Heston-Hull-White hybrid model.

\section{Appendices}
\appendix
\section{}
In order to obtain the dynamics for the SDEs in \eqref{Eq2.3} under ${\mathbb Q}^T$, we need to find the volatilities for both numeraires, respectively (refer to \cite{3}).
Denote the numeraire under $\mathbb Q$ as $N_{1,t} =
  e^{\int_{0}^{t}r(s)ds}$ and the
 numeraire under ${\mathbb Q}^{T}$ as $N_{2,t}=A(t,T)e^{-B(t,T)r(t)}$.
  Differentiating $\ln N_{1,t}$ yields
  \begin{equation*}
  \begin{array}{ll}
  \displaystyle
  d\ln N_{1,t}=r(t)dt=\left( \int_0^t\alpha^*(\beta^*-r(s)) \right) dt +\left(\int_{0}^t
  \eta \sqrt{r(s)}d\widetilde{W}_{3}(s)\right) dt,
  \end{array}
  \end{equation*}
  whereas the differentiation of $\ln N_{2,t}$ gives
  \begin{equation*}
  \begin{array}{ll}
  d\ln N_{2,t}= &\left(\dfrac{A'(t,T)}{A(t,T)} -B'(t,T)r(t) -B(t,T) \alpha^*(\beta^*-r(t))\right)dt \\
  &-B(t,T)
  \eta \sqrt{r(t)}d\widetilde{W}_{3}(t).
  \end{array}
  \end{equation*}
  Now we have obtained the volatilities for both numeraires as
  \begin{equation}
  \label{volatilities}
  \Sigma^{\mathbb Q}=\left(\begin{array}{ccc} 0 \\ 0 \\ 0
  \end{array} \right) \quad
  \mbox{and}  \quad
  \Sigma^{T}=\left(\begin{array}{ccc} 0 \\ 0 \\ -B(t,T)\eta\sqrt{r(t)}
  \end{array} \right).
  \end{equation}
Next,  the  drift term $\mu^{T}$  for the SDEs under $\mathbb{Q}^{T}$ is found by utilizing the formula below
  \begin{equation*}\label{ForwardMeasureSDE}
  \mu^{T}= \mu^{\mathbb Q} - \left(\Sigma \times L \times L^{T} \times (\Sigma^{\mathbb Q} -\Sigma^{T})\right),
  \end{equation*}
  with $\Sigma^{\mathbb Q}$ and $\Sigma^{T}$ in \eqref{volatilities}
  and the terms $\mu^{\mathbb Q}$, $\Sigma$ and $LL^{T}$ as defined in \eqref{Eq2.3}.
  \medskip
  This results in the transformation of \eqref{Eq2.3} under
  $\mathbb Q$ to the following system under the forward measure $\mathbb{Q}^{T}$
   \begin{eqnarray}
  \left( \begin{array}{cc} \frac {dS(t)}{S(t)} \\ d\nu(t)\\ dr(t)\\ \end{array} \right)
&  = & \Sigma \times L \times \left( \begin{array}{cc} dW^*_{1}(t)\\dW^*_{2}(t)\\dW^*_{3}(t)\end{array} \right)
  \\
&&+  \left( \begin{array}{cc} r(t)-\rho_{13}B(t,T)\eta\sqrt{r(t)}\sqrt{\nu(t)} \\ \kappa^*(\theta^*-\nu(t))-\rho_{23}\sigma B(t,T) \eta \sqrt{r(t)}\sqrt{\nu(t)}\\ \alpha^*\beta^*-(\alpha^*+B(t,T) \eta^2)r(t)\\ \end{array} \right) dt, \nonumber
  \end{eqnarray}
  where
  \begin{equation*}
  \begin{array}{ll}
  B(t,T)=\dfrac{2\left(e^{(T-t)\sqrt{(\alpha^*)^2+2\eta^2}}-1 \right)}{2\sqrt{(\alpha^*)^2 +2\eta^2}+
  \left(\alpha^*+\sqrt{(\alpha^*)^2+2\eta^2}\right)
  \left(e^{(T-t)\sqrt{(\alpha^*)^2+2\eta^2}}-1\right)}.
  \end{array}
  \end{equation*}



\section{}  

  \bigskip
The approximate solutions of the functions $E$ and $C$ can be found from the following differential equations which are obtained using the deterministic approximation technique discussed in \cite{12}
 \begin{equation*}
 \left \{
 \begin{array}{ll}
 \dfrac{dE}{d\tau}=\dfrac{1}{2} \eta^2 E^2 -(\alpha^*+B(T-\tau,T)\eta^2) E +\omega i, \\
 \dfrac{dC}{d\tau}=\kappa^*\theta^* D+ \alpha^* \beta^* E
 -\rho_{13}\eta \mathbb{E}^{T}\left[ \sqrt{\nu(T-\tau)}\sqrt{r(T-\tau)}\right]
 \omega i B(T-\tau,T)\\[0.8em]\quad \quad \quad+\rho_{13}\eta \mathbb{E}^{T}\left[\sqrt{\nu(T-\tau)}\sqrt{r(T-\tau)}\right]\omega i E b\\
 \quad \quad \quad -\rho_{23}\sigma\eta \mathbb{E}^{T}\left[\sqrt{\nu(T-\tau)}\sqrt{r(T-\tau)}\right]
 DB(T-\tau,T)\\[0.8em]\quad \quad \quad+\rho_{23}\eta \sigma \mathbb{E}^{T}\left[\sqrt{\nu(T-\tau)}\sqrt{r(T-\tau)}\right]DE,\\
 \end{array}
 \right.
 \end{equation*}
 with the initial conditions
 \begin{equation*}
 E(\omega,0)=0, \quad C(\omega,0)=0.
 \end{equation*}

  The differential equation related to $C$ 
  contains terms of $\sqrt{\nu(t)}\sqrt{r(t)}$ which are  non-affine. Note that standard techniques to find characteristic functions as in \cite{9} could not be applied in this case, thus
  we need to find approximations for these non-affine terms. The expectation $\mathbb{E}^{T}\left[\sqrt{\nu(t)}\right]$ with the CIR-type process
  can be approximated by, see \cite{12}:
   \begin{equation}\label{expecnu(t)ori}
  \mathbb{E}^{T}\left[\sqrt{\nu(t)}\right] \approx \sqrt{q_{1}(t)(\varphi_{1}(t)-1)+q_{1}(t)l_{1}+\dfrac{q_{1}(t)l_{1}}{2(l_{1}+\varphi_{1}(t))}}=:\Lambda_{1}(t),
  \end{equation}
  with
  \begin{equation}
  \label{q1t}
  q_{1}(t)=\dfrac{\sigma^{2}(1-e^{-\kappa^{*}t})}{4\kappa^*}, \quad l_{1}=\dfrac{4\kappa^{*}\theta^{*}}{\sigma^2}, \quad \varphi_{1}(t)=\dfrac{4\kappa^{*}\nu(0)e^{-\kappa^{*}t}}{\sigma^{2}(1-e^{-\kappa^{*}t})}.\\
  \end{equation}
  In order to avoid further complications during the derivation of the characteristic function and present a more efficient computation, the above approximation is further simplified as
  \begin{equation}\label{expecnu(t)simple}
  \mathbb{E}^{T}\left[\sqrt{\nu(t)}\right] \approx m_{1}+p_{1}e^{-Q_{1}t}=:\widetilde{\Lambda_{1}}(t),
  \end{equation}
  where
  \begin{equation}
  m_{1}=\sqrt{\theta^{*}-\frac{\sigma^2}{8\kappa^*}},  \quad
  \quad p_{1}=\sqrt{\nu(0)}-m_{1}, \quad Q_{1}=-\ln(p_{1}^{-1}(\Lambda_{1}(1)-m_{1})).\\
  \end{equation}

  The same procedure can be applied to find the expectation of $\mathbb{E}^{T}\left[\sqrt{r(t)}\right]$ as follows:
   \begin{equation}\label{expecnu(t)ori}
  \mathbb{E}^{T}\left[\sqrt{r(t)} \right] \approx \sqrt{q_{2}(t)(\varphi_{2}(t)-1)+q_{2}(t)l_{2}+\dfrac{q_{2}(t)l_{2}}{2(l_{2}+\varphi_{2}(t))}}=:\Lambda_{2}(t),\\
  \end{equation}
  with
  \begin{equation}
  \label{q2t}
  q_{2}(t)=\dfrac{\eta^{2}(1-e^{-\alpha^{*}t})}{4\alpha^*}, \quad l_{2}=\dfrac{4\alpha^{*}\beta^{*}}{\eta^2},
  \varphi_{2}(t)=\dfrac{4\alpha^{*} r(0)e^{-\alpha^{*}t}}{\eta^{2}(1-e^{-\alpha^{*}t})},\\
  \end{equation}
  and simplify further as
   \begin{equation}\label{expecr(t)simple}
  \mathbb{E}^{T}\left[\sqrt{r(t)}\right] \approx m_{2}+p_{2}e^{-Q_{2}t}=:\widetilde{\Lambda_{2}}(t),
  \end{equation}
  where
  \begin{equation}
  m_{2}=\sqrt{\beta^{*}-\frac{\eta^2}{8\alpha^*}},  \quad
  \quad p_{2}=\sqrt{r(0)}-m_{2}, \quad Q_{2}=-\ln(p_{2}^{-1}(\Lambda_{2}(1)-m_{2})).\\
  \end{equation}

  Utilizing the above expectations of both stochastic processes, we are able to obtain $\mathbb{E}^{T}\left[\sqrt{\nu(t)}\sqrt{r(t)}\right]$ by employing the following relation of dependent
  random variables and instantaneous correlation:
  \begin{equation*}
  \mathbb{E}^{T}\left[\sqrt{\nu(t)}\sqrt{r(t)}\right]=\mathbb{C}ov^{T}\left[\sqrt{\nu(t)},\sqrt{r(t)}\right] + \mathbb{E}^{T}\left[\sqrt{\nu(t)}\right]\mathbb{E}^{T}\left[\sqrt{r(t)}\right].
  \end{equation*}
  In order to figure out $\mathbb{C}ov^{T}\left[\sqrt{\nu(t)},\sqrt{r(t)}\right]$, we utilize the definition of instantaneous correlations:
  \begin{equation}\label{corr}
  {\rho}_{\mathsmaller{
  \sqrt{\nu(t)}\sqrt{r(t)}}}=\frac{\mathbb{C}ov^{T}\left[\sqrt{\nu(t)},\sqrt{r(t)}\right]}
  {\sqrt{\mathbb{V}ar^{T}\left[\sqrt{\nu(t)}\right]\mathbb{V}ar^{T}\left[\sqrt{r(t)}\right]}}.
  \end{equation}
  Substitution of the following
  \begin{equation*}
  \mathbb{V}ar^{T}\left[\sqrt{\nu(t)}\right]\approx
  \frac{\mathbb{V}ar^{T}\left[\nu(t)\right]}{4\mathbb{E}^{T}\left[\nu(t)\right]}
  \approx q_1(t)-\frac{q_1(t)l_{1}}{2(l_{1}+\varphi_{1}(t))}
  \end{equation*}
  and
  \begin{equation*}
  \mathbb{V}ar^{T}\left[\sqrt{r(t)}\right]\approx \frac{\mathbb{V}ar^{T}\left[r(t)\right]}
  {4\mathbb{E}^{T}\left[r(t)\right]}
  \approx q_2(t)-\frac{q_2(t)l_{2}}{2(l_{2}+\varphi_{2}(t))}
  \end{equation*}
  into \eqref{corr} gives us
  \begin{eqnarray*}
&&  \mathbb{C}ov^{T}\left[\sqrt{\nu(t)},\sqrt{r(t)}\right] \\
   \approx &&{\rho}_{\mathsmaller{ \sqrt{\nu(t)}\sqrt{r(t)}}}\left(\sqrt{\left(q_1(t)-\frac{q_1(t)l_{1}}{2(l_{1}+\varphi_{1}(t))} \right)\left(q_2(t)-\frac{q_2(t)l_{2}}{2(l_{2}+\varphi_{2}(t))}\right)}\right).
  \end{eqnarray*}


  \section{}  

In this appendix, we
derive approximate expressions of the
  expectations $\mathbb{E}_0^T\left[ e^{\widetilde{D}(\Delta t)\nu(t_{j-1})+\widetilde{E}(\Delta t)r(t_{j-1})}\right]$ and $\mathbb{E}_0^T\left[ e^{ \widehat{E}(\Delta t) r(t_{j-1})} \right]$.
 Then, we can obtain  an approximation for $G_j(\nu(0),r(0))$.

  The variables $\nu(t_{j-1})$ and $r(t_{j-1})$ can be  approximated by normally distributed random variables \cite{12} as follows:
  \begin{equation*}
  \nu(t)\approx \mathcal{N}\left(q_{1}(t)(l_{1}+\varphi_1(t)), {q_1(t)}^2(2l_{1}+4\varphi_{1}(t))\right),
  \end{equation*}
  and
  \begin{equation*}
  r(t)\approx \mathcal{N}\left(q_{2}(t)(l_{2}+\varphi_2(t)), {q_2(t)}^2(2l_{2}+4\varphi_{2}(t))\right),
  \end{equation*}
  where $q_1(t)$, $l_1$, $\varphi_1(t)$ are defined in \eqref{q1t} and  $q_2(t)$, $l_2$, $\varphi_2(t)$ are defined in \eqref{q2t}.
  Since both approximations of $\nu(t)$ and $r(t)$ are normally distributed, we can find the characteristic function of their sum which is also normally distributed.
  Let $Y(0,t_{j-1})=\widetilde{D}(\Delta t)\nu(t_{j-1})+\widetilde{E}(\Delta t)r(t_{j-1})$, then
  \begin{equation*}
  \mathbb{E}_0^T\left[ e^{Y(0,\;t_{j-1})}\right]\approx \exp\left(\mathbb{E}_0^{T}\left[Y(0,t_{j-1})\right]+\frac{1}{2} {\mathbb Var}^{T}\left[Y(0,t_{j-1})\right]\right),
  \end{equation*}
  where
    \begin{eqnarray*}
  && \mathbb{E}_0^{T}\left[Y(0,t_{j-1})\right] \\
  &\approx& \widetilde{D}(\Delta t)(q_{1}(t_{j-1})(l_{1}+\varphi_{1}(t_{j-1})))+\widetilde{E}(\Delta t)(q_{2}(t_{j-1})(l_{2}+\varphi_{2}(t_{j-1}))), \nonumber
  \end{eqnarray*}
  and
  \begin{eqnarray*}
&&  {\mathbb Var}^{T}\left[Y(0,t_{j-1})\right] \\
  &\approx &2\widetilde{D}(\Delta t)\widetilde{E}(\Delta t)\rho_{23}\sqrt{{q_1(t_{j-1})}^2(2l_{1}+4\varphi_1(t_{j-1}))}\sqrt{{q_2(t_{j-1})}^2(2l_{2}+4\varphi_{2}(t_{j-1}))}\\
  &&+ {\widetilde{D}(\Delta t)^2}(q_{1}(t_{j-1})^{2}(2l_{1}+4\varphi_{1}(t_{j-1}))) \\
  &&+ {\widetilde{E}(\Delta t)^2}(q_{2}(t_{j-1})^{2}(2l_{2}+4\varphi_{2}(t_{j-1}))).
  \end{eqnarray*}
  We can apply the same procedure to find the expression of $\mathbb{E}_0^T\left[ e^{ \widehat{E}(\Delta t) r(t_{j-1})} \right]$, which is given as follows:
  \begin{equation*}
  \begin{array}{ll}
  \mathbb{E}_0^T\left[ e^{ \widehat{E}(\Delta t) r(t_{j-1})} \right]\approx \exp \left(\mathbb{E}_0^{T}\left[\widehat{E}(\Delta t)r(t_{j-1})\right]
  +\frac{1}{2}{\mathbb Var}^{T}\left[\widehat{E}(\Delta t)r(t_{j-1})\right]\right) \\[0.8em]
  \quad\quad\quad\quad\quad\quad\quad\quad\approx \exp \left( \widehat{E}(\Delta t)
  (q_{2}(t_{j-1})(l_{2}+\varphi_{2}(t_{j-1}))) \right.\\[0.8em]
  \quad\quad\quad\quad\quad\quad\quad\quad\quad \left. + \frac{\widehat{E}(\Delta t)^2}{2} (q_{2}(t_{j-1})^{2}(2l_{2}+4\varphi_{2}(t_{j-1}))) \right).
  \end{array}
  \end{equation*}
Therefore, an approximation of  $G_j(\nu(0),r(0))$ is given as follows
  \begin{eqnarray*}
  &&G_j(\nu(0),r(0))\nonumber \\
\approx && e^{\widetilde{C}(\Delta t)} \cdot \exp \left( \widetilde{D}(\Delta t)(q_{1}(t_{j-1})(l_{1}+\varphi_{1}(t_{j-1})))
     +\widetilde{E}(\Delta t)(q_{2}(t_{j-1})(l_{2}+\varphi_{2}(t_{j-1}))) \right.
        \nonumber\\
&&         +\frac{\widetilde{E}(\Delta t)^2}{2}(q_{2}(t_{j-1})^{2}(2l_{2}+4\varphi_{2}(t_{j-1})))
   +\frac{\widetilde{D}(\Delta t)^2}{2}(q_{1}(t_{j-1})^{2}(2l_{1}+4\varphi_{1}(t_{j-1}))) \nonumber \\
&& \left. +\widetilde{D}(\Delta t)\widetilde{E}(\Delta t)\rho_{23}\sqrt{{q_1(t_{j-1})}^2(2l_{1}+4\varphi_1(t_{j-1}))}
\sqrt{{q_2(t_{j-1})}^2(2l_{2}+4\varphi_{2}(t_{j-1}))} \right) \nonumber\\
&&-2e^{\widehat{C}(\Delta t)}\cdot \exp \left( \widehat{E}(\Delta t) (q_{2}(t_{j-1})(l_{2}+\varphi_{2}(t_{j-1})) \right. \\
  &&\left. +\frac{\widehat{E}(\Delta t)^2}{2} (q_{2}(t_{j-1})^{2}(2l_{2}+4\varphi_{2}(t_{j-1})))\right) +1 \nonumber.
  \end{eqnarray*}

\end{document}